\documentstyle[aps,twocolumn,epsf,amssymb,floats]{revtex}

\newcommand{\PostScript}[3]{
  \vspace{#1cm}
\begin{center}
  \epsfysize=#2cm \leavevmode \epsfbox{#3}
\par
\end{center}
}

\begin{document}
\draft \preprint{preprint number}

\title{Dynamics of conversion of supercurrents into normal currents, and vice
versa}

\author{Arne Jacobs and Reiner K\"ummel}
\address{Institut f\"ur Theoretische Physik, Universit\"at W\"urzburg,
D-97074 W\"urzburg, Germany}

\date{\today} \maketitle
\begin{abstract}

The generation and destruction of the supercurrent in a superconductor ($S$)
between two resistive normal ($N$) current leads connected to a current source
is computed from the source equation for the supercurrent density.  This
equation relates the gradient of the pair potential's phase to electron and hole
wavepackets that create and destroy Cooper pairs in the $N/S$ interfaces.  Total
Andreev reflection and supercurrent transmission of electrons and holes are
coupled together by the phase rigidity of the non-bosonic Cooper-pair
condensate.  The calculations are illustrated by snapshots from a computer film.

\end{abstract}


\narrowtext \flushbottom

\section{Introduction}
Andreev scattering (AS) of electrons into holes and vice versa by spatial
variations of the superconducting pair potential \cite{and64}, in competition
and cooperation with conventional scattering, determines the electronic
structure and transport properties of inhomogeneous superconductors.  The
Tomasch effect in tunnel junctions \cite{mcm66}, Josephson currents
\cite{kul70,ish70,bar72,gun94,kro99,jac99}, excess currents, and subharmonic gap
structures \cite{blo82,oct83,kue90,gun94a} in superconducting ($S$)-normal
conducting ($N$)-superconducting junctions, as well as the transfer of half of
the Magnus force to the core electrons of a moving vortex line \cite{hof} are
due to AS.  It is involved in the persistent currents around the Aharonov-Bohm
flux in an $N/S$ metal loop \cite{bue86}, and there is AS in He$^3$, too
\cite{sch89}.  A wealth of AS phenomena has been discussed recently in
\cite{bag99} and \cite{oht99}.

While the conversion of a normal current into a supercurrent by electron
$\rightarrow$ hole scattering in the interface between an $N$ and an $S$ region
of semi-infinite lengths has been described before \cite{kue69,mat78,blo82}, the
reverse process has not.  Thus, it is the purpose of this paper to analyze the
normal-current $\longleftrightarrow$ supercurrent conversion processes in a
superconducting layer of finite length $L_z$ between two normal current leads.
These normal leads are connected to a reservoir (``battery'') which acts as the
current source in the closed circuit.  The extensions of the $N$ and $S$ regions
in $x$- and $y$-directions are $L_x$ and $L_y$; assuming that they don't exceed
the London penetration length one may neglect inhomogeneities of the current
density in the superconductor.  The metal-vacuum boundaries are treated as rigid
walls.  By varying $L_x$ and $L_y$ one can vary the dimensionality of the
system.  By showing in detail how in any transport experiment involving
superconductors electron $\leftrightarrow$ hole scattering brings about normal
current $\leftrightarrow$ supercurrent conversion our analysis may also prove
useful for the understanding of transport phenomena in quasi-two-dimensional
(Q2D) superconducting/semiconducting heterojunctions \cite{nit94,kroe94,bas98}
and superconducting quantum dots in Q2D channels.

\section{Charge conservation}

AS and the associated formation and destruction of Cooper pairs and
supercurrents can be calculated from the Time-dependent Bogoliubov-de Gennes
Equations (TdBdGE) \cite{kue69,mat78,blo82,kue99}.  They describe the evolution
of the spinor quasiparticle (q.p.)  wavefunction with the electron component
$u_{n}({\bf r},t)$ and the hole component $v_{n}({\bf r},t)$ under the influence
of scalar and vector potentials $V({\bf r},t)$ and ${\bf A}({\bf r},t)$ in the
single-electron Hamiltonian
\begin{displaymath}
H_0({\bf r},t) = \frac{1}{2m}\left[\frac{\hbar}{i}{\bf\nabla}-e{\bf A}({\bf r}
  ,t)\right]^2+V({\bf r},t)-\mu
\end{displaymath}
via the matrix equation
\begin{equation} \label{TdBdGE}
i\hbar \frac{\partial  }{\partial t}{u_{n}({\bf r},t) \choose v_{n}({\bf r},t)}
= \check{\mathcal{H}}({\bf r},t){u_{n}({\bf r},t) \choose v_{n}({\bf r},t)}.
\end{equation}
Here, the matrix Hamiltonian $\check{\mathcal{H}}({\bf r},t)$ has $H_0({\bf
r},t)$ and $-H_0^{*}({\bf r},t)$ in the diagonal, and the pair potential $\Delta
({\bf r},t)$ and its complex-conjugate in the off-diagonal.  The chemical
potential $\mu$ in $H_0$ is that of the reservoir.  We neglect all influences of
entropy production associated with current flow on the chemical potential,
because the number of degrees of freedom of the reservoir is assumed to be much
larger than that of the normal leads and the superconductor.  Thus, $\mu$ is
constant in space and time \cite{kue99}.  The index $n$ characterizes the
stationary q.p.  states from which the solutions of eq.\ (\ref{TdBdGE}) evolve
after time-dependent scalar and vector potentials have been switched on.

AS is a many-body process.  For its analysis it is convenient to consider a
non-equilibrium configuration $|T_{l\sigma}\rangle$ of the many-body system
where one quasiparticle state $(l\sigma)$, characterized by a tripel $l$ of
quantum numbers and spin $\sigma$, is definitely occupied and all other q.p.\
states $(n\sigma)$ are occupied according to the equilibrium distribution
function $f_n$.  All interactions that might affect the spin are neglected.
Then, it has been shown \cite{kue69,kue99} with the help of the TdBdGE
(\ref{TdBdGE}) that the expectation values $\langle T_{l\sigma}|\rho({\bf
r},t)|T_{l\sigma}\rangle$ and $\langle T_{l\sigma}|{\bf j}({\bf
r},t)|T_{l\sigma}\rangle$ of the many-body charge- and current-density operators
satisfy the relation
\begin{displaymath}
  \frac{\partial}{\partial t}\langle T_{l\sigma}|\rho({\bf
r},t)|T_{l\sigma}\rangle + {\rm div}\langle
T_{l\sigma}|{\bf j}({\bf r},t)|T_{l\sigma}\rangle
\end{displaymath}
\begin{displaymath}
  = -4\frac{e}{\hbar }{\rm Im}\left[\Delta^*({\bf r},t)u_l({\bf r},t)v_l^*({\bf
r},t)\right](1 - f_l) + {\rm div}{\bf j}_{sl}.
\end{displaymath}
The electron and hole wavefunctions $u_l$ and $v_l$ satisfy eq.\ (\ref{TdBdGE}),
and ${\bf j}_{sl}$ is the supercurrent density induced by the momentum and
charge transfer from the q.p.\ in $(l\sigma)$ to the superconducting condensate.
All mean-field q.p.\ states in ${\bf j}_{sl}$ and in the selfconsistency
equation for $\Delta ({\bf r},t)$ are in a Hilbert space specified to
$|T_{l\sigma}\rangle$.  Therefore, their wavefunctions all acquire the same
phase shift $S_l({\bf r},t)$ caused by the q.p.  in $(l,\sigma)$.  This leads to
a phase shift $2S_l({\bf r},t)$ of the pair potential, and ${\bf j}_{sl}$
becomes proportional to the gradient of $S_l({\bf r},t)$ times the number $N$ of
electrons in the superconductor \cite{kue69,kue99}.  (Recently the necessity of
a phase gradient for charge conservation in $N/S/N$ junctions has again been
pointed out by S\'anchez-Ca\~nizares and Sols \cite{san98}.)  The requirement
that charge is conserved in the many-body system results in the fundamental
source equation for the supercurrent density
\begin{equation} \label{qpsc}
 {\rm div}{\bf j}_{sl}
= 4\frac{e}{\hbar }{\rm Im}[\Delta^*({\bf r},t)u_l({\bf r},t)v_l^*({\bf r},t)]
(1 - f_l).
\end{equation}
This equation \cite{kue69,mat78,blo82,kue99} has a non-vanishing right-hand
side if the $u_l$ and $v_l$ describe quasiparticles that decay exponentially in
the superconductor during total Andreev reflection, because their energies are
less than the maximum value of $\Delta$.  In this case the source equation
yields a finite supercurrent ${\bf j}_{sl}$.  The phase shift of the pair
potential $2S_l$, on the other hand, is essentially given by the (integral of
the) r.h.s.  of eq.\ (\ref{qpsc}), divided by $N \ggg 1$ \cite{kue69}.  Thus,
despite of its importance for charge conservation, its numerically tiny value
can be neglected in our calculation of solutions of eq.\ (\ref{TdBdGE}).

\section{Current flow and representative wavepackets}

A shifted Fermi sphere (or its equivalent in Q2D and Q1D conductors) represents
the current-carrying many-body configurations in the two parts of the normal
current leads that are parallel to the $z$-axis and connected to the
superconductor.  (The direction of current flow in the parts bent towards the
reservoir is irrelevant.)  These leads, supposed to be much longer than the mean
inelastic scattering length, are conductors with resistance \cite{pot97}.  In
this non-equilibrium distribution of electrons above the Fermi surface in states
with positive momentum in $z$-direction and unoccupied states with negative
$z$-momentum below the Fermi surface the current-driving force from the battery
is balanced by the frictional forces from the energy-relaxation processes.  The
quasiparticles in this resistive-state configuration are uncorrelated.  Thus,
the total current in the closed circuit is the sum of the currents from the
individual quasiparticles.

In the following we try to obtain the details of normal current
$\longleftrightarrow$ supercurrent conversion by studying the motion of the
electrons (+) and holes (--) that are part of the shifted Fermi sphere.  Their
$z$-momenta are $\hbar k^{\pm}(E)$, with $ k^{\pm}(E) = [k_{zF}^2 \pm
E2m/\hbar^2]^{1/2}$.  Here, $k_{zF} =
[k_F^2-(n_x\pi/L_x)^2-(n_y\pi/L_y)^2]^{1/2}$ is the $z$-component of the Fermi
wavenumber $k_F \equiv [2m\mu/\hbar^2]^{1/2}$, and $(n_x\pi/L_x)$ and
$(n_y\pi/L_y)$, $n_{x,y}$ integers, are the wavenumbers of the standing waves
between the rigid walls that limit the metals in $x$- and $y$-directions.  The
energy $E$ of both electrons and holes is positive and measured relative to the
surface of the unshifted Fermi sphere at the chemical potential $\mu$.  For
normal current densities below the critical current densities of conventional
superconductors all $E$ are less than the modulus $\Delta$ of the pair potential
$\Delta(z) \approx \Delta\cdot\Theta(z)\Theta(L_z-z)$; here $\Theta(z)$ is the
Heavyside function which is sufficient to model the spatial variation of the
pair potential in the context of current flow
\cite{kul70,ish70,bar72,gun94,kro99,bue86,ave99,ben97}.  Details of the change
of $\Delta(z)$ close to the interfaces because of the proximity effect matter
little in the integral of eq.\ (\ref{qpsc}) that yields the supercurrent.

The motion of wavepackets shows best the dynamics of quantum-mechanical
processes.  Thus, similar to the use of representative, well-controlled,
preformed wavepackets in the calculation of differential cross sections for
conventional scattering, we investigate the current dynamics in a closed $N/S/N$
circuit by studying the motion of wavepackets that are representative for the
electronic configuration in any transport experiment involving a superconductor
connected to a current source by normal conductors.  Conventional scattering
processes are disregarded because we are only interested in AS in the $N/S$
interfaces.  In principle, impurity scattering could be treated with the help of
the scattering matrix formulation \cite{ben97,joh99}.  This is especially
convenient for devices that, unlike the ones considered by us, involve only a
small number of incoming and outgoing channels.  Interface barriers that weaken
Andreev scattering have been considered previously \cite{blo82,oct83,schue93};
the competition between conventional and Andreev scattering has been discussed
in terms of the diagonal and off-diagonal forces associated with broken
symmetries \cite{gun94}, and a motion picture of the wavepacket dynamics in such
a case can be viewed in the Internet under the address given in the caption of
Fig.\ 1.  If conventional scattering is present, the Gaussian spectral function
$D(E)$, see below, has to be multiplied by the probability amplitude of
transmission.  There are only few scattering impurities and no interface
barriers present in a Q2D electron gas in a modulation-doped system consisting
of an InAs channel between an AlSb substrate and a superconducting Niobium layer
that induces a pair potential in the electron gas via the proximity effect
\cite{pp}.  For such an experimental set-up our calculations apply exactly
(within the limits of the Andreev approximation).  For the general case of any
superconductor between any two normal current leads they show the quantum
mechanically and electrodynamically essential dynamics of AS that rules the
charge transport in addition to conventional scattering.

We start our analysis with the initial condition that a normalized electron
wavepacket, localized around $z_0 < 0$ in the normal current lead to the left of
the superconductor at time $t=0$, travels towards the superconductor.  In the
center of the wavepacket the energy is $E_l < \Delta$.  (By varying $E_l$ and
$k_{zF}$ one can obtain all the low-lying electron excitations that are part of
the shifted Fermi sphere in the left normal lead.)  For convenience we choose a
Gaussian spectral function
\begin{displaymath} D(E) =
 \frac{a_z}{\sqrt{2\pi}}\,e^{-[k^+(E)-k^+(E_l)]^2a_z^2/2}e^{-i[k^+(E)-k^+(
 E_l)]z_0};
\end{displaymath}
the position-uncertainty parameter $a_z \ll |z_0|$ is chosen so large that the
related energy spread of the wavepacket, $\delta_E = \hbar^2 k^+(E_l)/ma_z$, is
less than $\Delta - E_l$.  Solutions of eq.\ (\ref{TdBdGE}) are calculated
neglecting $V$ and ${\bf A}$ and approximating $\Delta({\bf r},t)$ by the real
$\Delta(z)$, thereby disregarding repercussions of the quasiparticle-induced
supercurrent on the q.p.\ \cite{hof} and on itself.  (As discussed above, the
phase shift of the pair potential due to one Andreev reflection process is
negligibly small.)  These solutions are multiplied by the spectral function
$D(E)$, integrated over all energies, and matched at the left $N/S$ interface,
i.e.  $z=0$, in the usual Andreev approximation where terms of the order of
$\Delta/\mu$ are neglected outside the exponentials.  Energy-dependent functions
are Taylor expanded around $E_l$ up to first order in $(E-E_l)$.  This affects
the amplitude of the Andreev-reflection probability
\begin{displaymath}
  \gamma(E) \equiv e^{-i\arccos E/\Delta} \approx \gamma(E_l)e^{(i/\hbar)
  (E-E_l)\tau_l}
\end{displaymath}
and $k^{\pm}(E) = [k_{zF}^2 \pm E2m/\hbar^2]^{1/2} \approx k_l^{\pm} \pm
(E-E_l)/\hbar v_l^{\pm} $; \- $\tau_l = \hbar[\Delta^2-E_l^2]^{-1/2}$ is the
time for one electron $\rightarrow$ hole-scattering event and the associated
formation of a Cooper pair, see eqs.\ (\ref{unl},\ref{vnl}), and
\begin{center} 
$k_l^{\pm}\equiv k^{\pm}(E_l)$, \quad $v_l^{\pm} \equiv \hbar k_l^{\pm}/m.$
\end{center} 
The resulting electron and hole wavepackets $u_{NL}({\bf r}, t)$ and
$v_{NL}({\bf r},t)$ in the left normal current lead, $z<0$, and the
exponentially decaying solutions $u_{SL}({\bf r},t)$ and $v_{SL}({\bf r},t) $ in
$z>0$, that contribute to the source equation (\ref{qpsc}) essentially in the
left half of the superconductor, turn out to be
\begin{eqnarray}
u_{NL}& = & w_le^{ik_l^+z}e^{-[z_0-z+v_l^+t]^2/2a_z^2},\label{unl}\\
v_{NL}& = & w_l\gamma(E_l)e^{ik_l^-z}
e^{-[z_0+(v_l^+/v_l^-)z+v_l^+(t-\tau_l)]^2/2a_z^2},\label{vnl}\\
u_{SL}& = & w_le^{ik_{zF}z}e^{-\kappa_l z}e^{-[z_0+v_l^+t]^2/2a_z^2},
\label{us} \\
v_{SL}& = & w_l\gamma(E_l)e^{ik_{zF}z}
e^{-\kappa_l z}e^{-[z_0+v_l^+(t-\tau_l)]^2/2a_z^2},\label{vs}
\end{eqnarray}
where
\[
w_l\equiv \frac{2}{(L_xL_ya_z\sqrt{\pi})^{1/2}}\sin\left(\frac{n_x\pi}{L_x}x
\right)\sin\left(\frac{n_y\pi}{L_y}y\right)e^{-iE_lt/\hbar},
\]
and $\kappa_l \equiv (\Delta^2-E_l^2)^{1/2}/\hbar v_{zF} = 1/\tau_l v_{zF}$,
with $v_{zF} \equiv \hbar k_{zF}/m$. (For the sake of simplicity the
complex wavenumbers in $u_{SL}$ and $v_{SL}$ have not been Taylor expanded
in $(E-E_l)$ but rather taken at $E_l$ right away.)

Identifying the wavefunctions $u_{SL}$ and $v_{SL}$ from eqs.\ (\ref{us})
and (\ref{vs}) with the $u_l$ and $v_l$ of the source equation (\ref{qpsc})
and integrating that equation from $z=0$ to $z$ yields the density of the
supercurrent in $z$-direction, induced in the left half of the
superconductor by AS of the electron wavepacket $u_{NL}$ into the hole
wavepacket $v_{NL}$, as
\begin{eqnarray} \label{js}
{\bf j}_{sl,L} &=& {\bf e}_z(2ev_{zF})|w_l|^2[1- e^{-2\kappa_l z}]\nonumber\\
&&\times e^{-\{[z_0+v_l^+(t-\tau_l/2)]^2 + (\tau_l v_l^+/2)^2\} /a_z^2}(1-f_l).
\end{eqnarray}
Here we have assumed that $L_z \gg 1/\kappa_l$.  In the opposite case one would
have to add a second source term on the r.h.s.  of eq.\ (\ref{qpsc}).  This term
would contain the contribution from the wavepacket solution $u_{SR}({\bf r},t),
v_{SR}({\bf r},t)$ of eq.\ (\ref{TdBdGE}) for $0<z<L_z$ that matches to the
current-carrying q.p.\ wavepackets in the right normal current lead at $z=L_z$
and decays exponentially with increasing distance from $L_z$.  However, in our
case of large $L_z$ the $u_{SR}({\bf r},t)$ and $v_{SR}({\bf r},t)$ give only
rise to the supercurrent ${\bf j}_{sl,R}$ in the right half of the
superconductor.  The complex amplitudes of the solutions of the TdBdGE in the
energy integrals that form the wavepackets $u_{SR}({\bf r},t)$ and $v_{SR}({\bf
r},t)$ are uniquely determined by the requirement that the supercurrent
densities ${\bf j}_{sl,L}$ and ${\bf j}_{sl,R}$, computed from $u_{SL}v_{SL}^*$
and $u_{SR}v_{SR}^*$, join smoothly at all times somewhere within the
superconductor.  Since the phases of the Fermi-liquid quasiparticles in the left
and right normal current leads are at random, only the current densities, not
the wavefunctions, must join smoothly.  (If, on the other hand, the
superconductor were only a thin slab with $L_z \ll 1/\kappa_l$, there would be a
finite probability that the q.p.  do not suffer AS and induce a supercurrent but
rather carry their phases through the pair-potential wall in a tunneling
process.)  The matching point turns out to be $L_z/2$, and the evanescent
wavepackets result to be
\begin{eqnarray}
u_{SR}& = & w_le^{ik_{zF}z}e^{-\kappa_l(L_z-z)}
e^{-[z_0+v_l^+(t-\tau_l)]^2/2a_z^2},\label{ur} \\
v_{SR}& = & w_l\gamma(E_l)^{-1}e^{ik_{zF}z}e^{-\kappa_l(L_z-z)}
e^{-[z_0+v_l^+t]^2/2a_z^2}. \label{vr}
\end{eqnarray}
More details are given in the Appendix.

The supercurrent density ${\bf j}_{sl,R}$, obtained by integrating eq.\
(\ref{qpsc}) from $L_z$ to $z$, with $u_{SR}v_{SR}^*$ in the place of
$u_lv_l^*$, has the same form as ${\bf j}_{sl,L}$ of eq.\ (\ref{js}) except that
$\exp[-2\kappa_l z]$ is being replaced by $\exp[-2\kappa_l(L_z-z)]$.  Finally,
the wavepacket solutions $u_{NR}({\bf r},t), v_{NR}({\bf r},t)$ of eq.\
(\ref{TdBdGE}) in the right normal current lead, $z>L_z$, that match to the
$u_{SR}({\bf r},t), v_{SR}({\bf r},t)$ at the right interface in $z=L_z$, become
\begin{eqnarray} \label{unr}
u_{NR}& = & w_le^{ik_{zF}L_z}e^{ik_l^+(z-L_z)}
\nonumber\\&&\times
e^{-[z_0+L_z-z+v_l^+(t-\tau_l)]^2/2a_z^2},
\end{eqnarray}
\begin{eqnarray}\label{vnr}
v_{NR}& = & w_l\gamma(E_l)^{-1}e^{ik_{zF}L_z}e^{ik_l^-(z-L_z)}
\nonumber\\&&\times
e^{-[z_0+(v_l^+/v_l^-)(z-L_z)+v_l^+t]^2/2a_z^2}.
\end{eqnarray}
Note that these wavepackets are the result of only {\em one} initial condition,
namely ``electron wavepacket incident from the left''.  This initial condition,
the matching of the wavepackets at the interfaces, the requirement of charge
conservation as expressed by eq.\ (\ref{qpsc}), and the smooth joining of the
supercurrent densities ${\bf j}_{sl,L}$ and ${\bf j}_{sl,R}$ determine
unequivocally the wavepackets in the right normal current lead (apart from an
irrelevant, constant phase factor that has deliberately been set equal to
unity).

\section{Andreev scattering and Cooper pair formation}

Comparison of the $u_{NL}, v_{NL}$ from eqs.\ (\ref{unl},\ref{vnl}) with the
$u_{NR}, v_{NR}$ from eqs.\ (\ref{unr},\ref{vnr}) shows that the center of the
electron wavepacket $u_{NL}$, propagating to the right with velocity $v_l^+$ in
the left normal current lead, and the center of the hole wavepacket $v_{NR}$,
propagating to the left with velocity $v_l^-$ in the right normal current lead,
hit the left and right interfaces at $z=0$ and $z=L_z$ (with different phases)
at the same time $t_0 = -z_0/v_l^+$, while the hole wavepacket $v_{NL}$,
propagating to the left in $z<0$ with $v_l^-$, and the electron wavepacket
$u_{NR}$, propagating to the right in $z>L_z$, are retarded by the time $\tau_l$
with respect to the incident wavepackets.  The holes moving to the left
transport positive momentum and negative charge to the right, just as the
electrons of opposite group velocity do.  The supercurrent density ${\bf
j}_{sl}$ spreads ``instantaneously'' throughout the superconductor (the velocity
of light not exceeding, of course) and couples directly electron $\rightarrow$
hole scattering in the left to hole $\rightarrow$ electron scattering in the
right interface.

This ``instantaneous'' coupling may seem to be surprising, but only at a first
look.  Giving it a second thought one sees that our result confirms by explicit
wavepacket calculation, and for the first time, as far as we know, what one has
concluded before intuitively and from steady state calculations:  hole
$\rightarrow$ electron scattering destroys and electron $\rightarrow$ hole
scattering creates supercurrents, whenever currents flow through $S/N$ and $N/S$
interfaces, e.g.  in vortex lines \cite{hof}.  Both scattering processes must
occur simultaneously even at far-apart interfaces because of charge conservation
in closed circuits {\em and} as a direct consequence of the very essence of BCS
superconductivity:  Cooper pairs are {\bf not} bosons, despite of what one can
read frequently, because within the volume of one Cooper pair of a conventional
superconductor there are about one million of mass centers of other Cooper
pairs, their creation and destruction operators do not satisfy bosonic
commutation relations, and their condensate wavefunction is antisymmetric
\cite{sch64}.  Therefore, the Cooper pair formed by total Andreev reflection in
one interface cannot exist outside but only within the condensate to which the
momentum 2$\hbar k_{zF}$ and the charge of the two electrons of opposite spin,
that have entered the superconductor, must be transferred.  Because of the
``phase rigidity of the electron pair fluid'' \cite{and84}, so typical for
off-diagonal long range order, this charge and momentum transfer, and the
related phase shift, manifest themselves immediately in a current flow out of
the other interface.  Thus, no charges can accumulate in the superconductor
according to its capacitance (as charges from quasiparticles with energies $E >
\Delta$ may do), and the reservoir has to emit holes into and receive electrons
from the right normal lead at the same rate at which it emits electrons into and
receives holes from the left current lead.  This is illustrated by the snapshots
from a computer film in Figure 1.  The Andreev-reflected wavepackets may be
considered as supercurrent-transmitted wavepackets as well.  Together with the
incident wavepackets they represent any of the low lying single-particle
excitations with energies $E_l < \Delta$ in the current-carrying, resistive
parts of the circuit.

Considering the supercurrent contribution from our representative
quasiparticle-wavepacket configuration we note that its spatial maximum is
at $z=L_z/2$, and its maximum in time occurs at $t = t_0 + \tau_l/2$. From
the equations for ${\bf j}_{sl}$, $u_{SL,R}$ and $v_{SL,R}$ one sees that
the quasiparticle current density
\begin{displaymath} {\bf j}_{QPl} \equiv
  \frac{e}{m}{\rm Re}\left[u_l^*\frac{\hbar}{i}{\bf \nabla}u_l -
  v_l\frac{\hbar}{i}{\bf \nabla}v_l^*\right](1-f_l)
\end{displaymath}
changes into the supercurrent density ${\bf j}_{sl}$ and vice versa within
a distance $1/\kappa_l = v_{zF}\tau_l$ from the interfaces.

In conclusion, the combinations of single-particle excitations (wavepackets) and
collective modes (supercurrents), connected by AS as shown in Fig.\ 1, are the
consequence of the phase rigidity of the Cooper-pair condensate and the
adjustment of the current configuration to charge conservation in the $N/S$ and
$S/N$ interfaces.  They are the current-carrying elementary excitations in
closed $N/S/N$ circuits.

\section{Outlook}

The supercurrent, carried by the condensate in the $S$ layer, involves only
states with $|E_n| \geq \Delta$.  It continues the current in the normal current
leads where all electrons and holes have energies $E_l < \Delta$.  If the total
current density exceeds its critical value, i.e.\ if the center of the Fermi
sphere in the normal current leads is shifted by more than $\hbar q_{cS}= \Delta
m/\hbar k_F$, depairing sets in, and when superconductivity has broken down, the
uncorrelated normal-state configuration reigns everywhere in the circuit.  If,
on the other hand, the single $S$ layer is replaced by a mesoscopic $SNS$
junction, the many-body configuration in the central $N$ layer is a
phase-coherent one and thus different from the uncorrelated configurations in
the normal current leads.  In an $N/SNS/N$ circuit the $SNS$ junction acts as a
gapless superconductor \cite{ish70}.  It can carry a dissipation-free Josephson
current through the central $N$ layer via phasecoherent q.p.\ states with
$|E|<\Delta$ and $|E| \geq \Delta$ \cite{gun94}.  This current converts as a
whole into the total supercurrent of the $S$ layers, and vice versa, whereas,
according to eq.\ (\ref{qpsc}), each uncorrelated q.p.\ from the normal current
leads individually induces its proper fraction of the total supercurrent.  If
the total current density exceeds the critical Josephson-current density at a
Fermi-sphere shift of $\hbar q_{cSNS}\approx\hbar/d$, where $d$ is the length of
the central $N$ layer \cite{bar72}, a voltage drop appears across the $SNS$
junction.  There are different models \cite{oct83,kue90,gun94a,ave95,jac99} for
$SNS$ junctions with voltage drops.  They differ with respect to the implicit
assumptions about the rate and energy range of q.p.\ creation in the central $N$
layer by supercurrent $\longrightarrow$ quasiparticle-current conversion.  The
question of how this rate and range depend upon the weakening of phase coherence
in the $SNS$ junction by energy exchange between quasiparticles and electric
fields, phonons, and thermal fluctuations like Nyquist-Johnson noise
\cite{nin84} is presently investigated.\\

{\bf  APPENDIX: MATCHING OF SUPERCURRENTS}

\setcounter{equation}{0}
\renewcommand{\theequation}{\mbox{A\arabic{equation}}}

The supercurrent density in the right part of the 
superconductor, $j_{sl,R}$, results from the evanescent wavepackets 
$u_{SR}, v_{SR}$.
These are built up from the  stationary solutions of eq.\ (\ref{TdBdGE}) that
decay exponentially within the superconductor with increasing distance from 
the right interface at 
$z=L_z$. The free amplitudes $\Omega(E)$ of each of these solutions are
determined by demanding 
that $j_{sl,R}$ joins smoothly with the supercurrent density $j_{sl,L}$
from the left part of  the superconductor  at all times somewhere 
within the superconductor. Once the amplitudes $\Omega(E)$ are known,  
the electron- and hole-wavepackets  in the right normal current lead are
unequivocally determined by eq.\ (\ref{TdBdGE}) and the matching conditions
for the wavepackets in $z=L_z$. 

The remainder of this appendix just shows how the amplitudes  $\Omega(E)$ are
calculated.

In the right part of the superconductor,  with the Gaussian spectral 
function $D(E)$ and those solutions of eq.\ (\ref{TdBdGE}) that increase as 
$z<L_z$ approaches $L_z$, we obtain
\begin{eqnarray}
u_{SR}&=&\sin\left(\frac{n_x\pi}{L_x}x\right)\sin\left(\frac{n_y\pi}{L_y}y\right)
\frac{1}{\delta_E\sqrt{2\pi}}\;e^{-iE_lt/\hbar}e^{ik_{zF}z}e^{-\kappa_l(L_z-z)}
\nonumber\\
&&\times\int_0^\infty\mbox{d}Ee^{-\left[\left(a_z/\hbar v_l^+\right)^2(E-E_l)^2/2
+i\left(z_0/\hbar v_l^++t/\hbar\right)(E-E_l)\right]}\Omega(E),\label{usrint}\\
v_{SR}&=&\sin\left(\frac{n_x\pi}{L_x}x\right)\sin\left(\frac{n_y\pi}{L_y}y\right)
\frac{1}{\delta_E\sqrt{2\pi}}\;e^{-iE_lt/\hbar}e^{ik_{zF}z}e^{-\kappa_l(L_z-z)}
\gamma(E_l)^{-1}\nonumber\\
&&\times\int_0^\infty\mbox{d}Ee^{-\left[\left(a_z/\hbar v_l^+\right)^2(E-E_l)^2/2
+i\left(z_0/\hbar v_l^++(t+\tau_l)/\hbar\right)(E-E_l)\right]}\Omega(E).
\label{vsrint}
\end{eqnarray}
(The term $\delta_E = \hbar^2k^+(E_l)/ma_z $  in the 
denominator, which results 
from replacing the wavepacket integration over $k^+$ by one over $E$, drops out
after the evaluation of the integrals.)

Let $z^\prime$ be the point where the supercurrent densities ${\bf j}_{sl,L}$
and ${\bf j}_{sl,R}$ join smoothly.  Then the integrals of the source equation
(\ref{qpsc}) must satisfy
\clearpage
\begin{eqnarray}
{\bf j}_{sl,L}\Big|_{z^\prime}&=&{\bf e}_z\int_0^{z^\prime}\mbox{d}z\;4
\frac{e}{\hbar}\mbox{Im}\left(\Delta^*u_{SL}v_{SL}^*\right)(1-f_l)\nonumber\\
&\stackrel{!}{=}&-{\bf e}_z\int_{z^\prime}^{L_z}
\mbox{d}z\;4\frac{e}{\hbar}\mbox{Im}\left(\Delta^*u_{SR}v_{SR}^*\right)(1-f_l)=
{\bf j}_{sl,R}\Big|_{z^\prime}.\label{jgleichj}
\end{eqnarray}

For convenience we write the complex  $\Omega(E)$ as the product of 
two factors, one of them being energy-dependent:
\begin{equation}
\Omega(E)\equiv\omega_1\omega_2(E).
\end{equation}

Now we insert the $u_{SR}$ and 
$v_{SR}$ of eqs.\ (\ref{usrint}) and (\ref{vsrint}) into the supercurrent 
density ${\bf j}_{sl,R}|_{z^\prime}$ of eq.\ (\ref{jgleichj})  and, similarly, 
${\bf j}_{sl,L}|_{z^\prime}$ 
is expressed by the solutions $u_{SL}$ and $v_{SL}$ in the same form, i.e. 
without evaluating the energy integrals: 
\begin{eqnarray}
{\bf j}_{sl,L}\Big|_{z^\prime}&=&{\bf e}_z4\frac{e}{\hbar}\Delta\sin^2\left(
\frac{n_x\pi}{L_x}x\right)\sin^2\left(\frac{n_y\pi}{L_y}y\right)
\frac{1}{2\pi\delta_E^2}\frac{4}{L_xL_ya_z\sqrt{\pi}}\nonumber\\
&&\times\mbox{Im}\left\{\gamma(E_l)^*\left[\int_0^\infty\mbox{d}Ee^{-\left[\left(
a_z/\hbar v_l^+\right)^2(E-E_l)^2/2+i\left(z_0/\hbar v_l^++t/\hbar\right)(E-E_l)
\right]}\right]\right.\nonumber\\
&&\times\left.\left[\int_0^\infty\mbox{d}Ee^{-\left[\left(a_z/\hbar v_l^+\right)^2
(E-E_l)^2/2+i\left(z_0/\hbar v_l^++(t-\tau_l)/\hbar\right)(E-E_l)\right]}\right]^*
\right\}\nonumber\\
&&\times\int_0^{z^\prime}\mbox{d}ze^{-2\kappa_l z}(1-f_l)\label{jlint},\\
{\bf j}_{sl,R}\Big|_{z^\prime}&=&-{\bf e}_z4\frac{e}{\hbar}\Delta\sin^2\left(
\frac{n_x\pi}{L_x}x\right)\sin^2\left(\frac{n_y\pi}{L_y}y\right)\frac{1}{2\pi
\delta_E^2}|\omega_1|^2\nonumber\\
&&\times\mbox{Im}\left\{\left[\gamma(E_l)^{-1}\right]^*\left[\int_0^\infty\mbox{d}E
e^{-\left[\left(a_z/\hbar v_l^+\right)^2(E-E_l)^2/2+i\left(z_0/\hbar v_l^++t/\hbar
\right)(E-E_l)\right]}\omega_2(E)\right]\right.\nonumber\\
&&\times\left.\left[\int_0^\infty\mbox{d}Ee^{-\left[\left(a_z/\hbar v_l^+\right)^2
(E-E_l)^2/2+i\left(z_0/\hbar v_l^++(t+\tau_l)/\hbar\right)(E-E_l)\right]}\omega_2(E)
\right]^*\right\}\nonumber\\
&&\times\int_{z^\prime}^{L_z}\mbox{d}ze^{-2\kappa_l(L_z-z)}(1-f_l)\label{jrint}.
\end{eqnarray}
Note that
\begin{equation}
\mbox{Im}\left\{\left[\gamma(E_l)^{-1}\right]^*\right\}=-\mbox{Im}\left\{
\gamma(E_l)^*\right\}.
\end{equation}

We demand that the integrals over $z$ are equal at $z^\prime$:
\begin{equation}
\int_0^{z^\prime}\mbox{d}ze^{-2\kappa_l z}=\frac{1}{2\kappa_l}\left(1-e^{-2\kappa_l 
z^\prime}\right)\stackrel{!}{=}
\frac{e^{-2\kappa_lL_z}}{2\kappa_l}\left(e^{2\kappa_l L_z}-e^{2\kappa_l z^\prime}
\right)=\int_{z^\prime}^{L_z}\mbox{d}ze^{-2\kappa_l(L_z-z)}.
\end{equation}
This equation is satisfied by
\begin{equation}
z^\prime=\frac{1}{2}L_z\label{o3}.
\end{equation}
The  energy integrals in eqs.\ (\ref{jlint}) and (\ref{jrint}) are real. 
By comparing them, i.e. the first one in eq.\ (\ref{jlint}) with the second one
in eq.\ (\ref{jrint}), or  the second one in (\ref{jlint}) with the first one 
in eq.\ (\ref{jrint}), we find
\begin{equation}
\omega_2(E)=e^{i(E-E_l)\tau_l/\hbar}.
\end{equation}
Finally, comparison of  the prefactors in eqs.\ (\ref{jlint}) and 
(\ref{jrint}) yields
\begin{equation}
\omega_1=\frac{2}{(L_xL_ya_z\sqrt{\pi})^{1/2}}.
\end{equation}
Thus, the complex amplitudes of the solutions in the right part of the 
superconductor are given by
\begin{equation}
\Omega(E)=\frac{2}{(L_xL_ya_z\sqrt{\pi})^{1/2}}e^{i(E-E_l)\tau_l/\hbar}.
\end{equation}

\begin{figure}[htb]
\PostScript{0}{8.8}{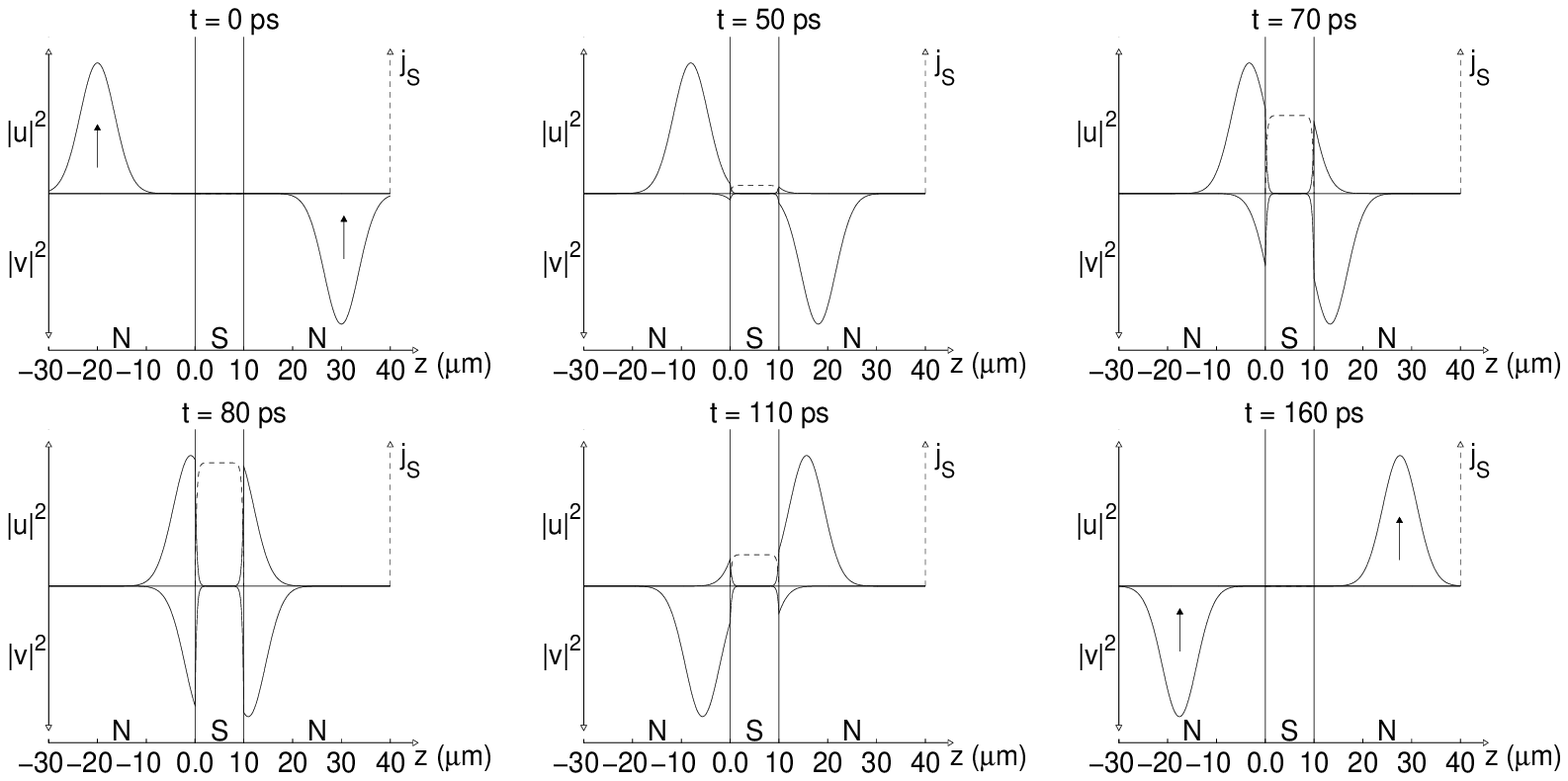}
\caption{Propagation and Andreev scattering of the probability densities
$|u|^2$ and $|v|^2$ of  a representative spin-up electron- and
hole-wavepacket configuration (solid lines), and the induced supercurrrent
density ${\bf j}_s$ (dashed line), in a current-carrying closed $N/S/N$
circuit.  For the sake of clear graphical representation the
initial conditions for the electron wavepacket, incident from the left, have
been chosen as:
energy in the center $E_l = \Delta/2 = 0.15 \mbox{ meV}$, spatial spread
$a_z = 5 \mbox{ $\mu$m}$, and $k_{zF} = 0.9k_F = 2.06 \mbox{ nm}^{-1}$.
Via charge conservation by supercurrent induction these conditions determine
the parameters and the motion of the
resulting hole wavepacket incident from the right; the retardation time for
AS,  $\tau_l = \hbar/[\Delta^2-E_l^2]^{1/2}$, is about two picoseconds (ps).
 More  computer films on electron $\rightarrow$ hole
and electron $\rightarrow$ electron scattering in one $N/S$ interface, also
for energies above the gap,  can be viewed in the Internet under {\em
http://theorie.physik.uni-wuerzburg.de/TP1/kuemmel/films/filmse.html}.}

\label{label1}
\end{figure}


\end{document}